\documentclass[twocolumn,PRB,aps,psfig,preprintnumbers,superscriptaddress]{revtex4}
\usepackage{graphicx}
\usepackage{epstopdf}
\usepackage{float}
\usepackage{color}
\usepackage{amsmath}
\usepackage{bm}

\begin{document}

\title{Microscopic characterization of the magnetic properties of the itinerant antiferromagnet  La$_2$Ni$_7$  by $^{139}$La NMR/NQR measurements }

\author{Q.-P. Ding}
\affiliation{Ames National Laboratory, U.S. DOE, Ames, Iowa 50011, USA}
\author{J. Babu}
\affiliation{Ames National Laboratory, U.S. DOE, Ames, Iowa 50011, USA}
\affiliation{Department of Physics and Astronomy, Iowa State University, Ames, Iowa 50011, USA}
 \author{K. Rana}
 \affiliation{Ames National Laboratory, U.S. DOE, Ames, Iowa 50011, USA}
\author{Y. Lee}
\affiliation{Ames National Laboratory, U.S. DOE, Ames, Iowa 50011, USA}
\author{S. L. Bud'ko}
\affiliation{Ames National Laboratory, U.S. DOE, Ames, Iowa 50011, USA}
\affiliation{Department of Physics and Astronomy, Iowa State University, Ames, Iowa 50011, USA}
\author{R. A. Ribeiro}
\affiliation{Ames National Laboratory, U.S. DOE, Ames, Iowa 50011, USA}
\affiliation{Department of Physics and Astronomy, Iowa State University, Ames, Iowa 50011, USA}
\author{P. C. Canfield}
\affiliation{Ames National Laboratory, U.S. DOE, Ames, Iowa 50011, USA}
\affiliation{Department of Physics and Astronomy, Iowa State University, Ames, Iowa 50011, USA}
\author{Y. Furukawa}
\affiliation{Ames National Laboratory, U.S. DOE, Ames, Iowa 50011, USA}
\affiliation{Department of Physics and Astronomy, Iowa State University, Ames, Iowa 50011, USA}

\date{\today}

\begin{abstract} 
   $^{139}$La nuclear magnetic resonance (NMR) and nuclear quadrupole resonance (NQR) measurements have been performed to investigate the magnetic properties of the itinerant magnet La$_2$Ni$_7$ which shows a series of antiferromagnetic (AFM) phase transitions at $T_{\rm N1}$ = 61 K, $T_{\rm N2}$ = 56 K, and $T_{\rm N3}$~=~42 K under zero magnetic field.
   Two distinct La NMR signals were observed due to the two crystallographically inequivalent La sites in La$_2$Ni$_7$ (La1 and La2 in the La$_2$Ni$_4$ and the LaNi$_5$ sub-units of the La$_2$Ni$_7$ unit cell, respectively).
   From the $^{139}$La NQR spectrum in the AFM state below $T_{\rm N3}$, the AFM state was revealed to be a commensurate state where Ni ordered moments align along the crystalline $c$ axis.
    Owing to the two different La sites, we were able to estimate the average values of the Ni ordered moments (0.09-0.10 $\mu_{\rm B}$/Ni and 0.17$\mu_{\rm B}$/Ni around La1 and La2, respectively)  from $^{139}$La NMR spectrum measurements in the AFM state below $T_{\rm N3}$, suggesting a non-uniform distribution of the Ni-ordered moments in the AFM state. 
     In contrast, a more uniform distribution of the Ni-ordered moments in the saturated paramagnetic state induced by the application of  high magnetic fields is observed.
    The temperature dependence of the sublattice magnetization measured by the internal magnetic induction at the La2 site in the AFM state was reproduced by a local-moment model better than the self-consistent renormalization (SCR) theory for weak itinerant antiferromagnets.
   Given the small Ni-ordered moments in the magnetically ordered state, our results suggest that La$_2$Ni$_7$ has characteristics of both itinerant and localized natures in its magnetism.
   With this in mind, it is noteworthy that the temperature dependence of nuclear spin-relaxation rates (1/$T_1$) in the paramagnetic state above $T_{\rm N1}$ measured at zero magnetic field can be explained qualitatively by both the SCR theory and the local-moment model.

\end{abstract}

\maketitle

  \section{ Introduction}

    Fragile magnets with small ordered moments, such as weak itinerant ferromagnets and antiferromagnets, have attracted much interest in studying quantum phase transition since the magnetic states of those compounds can be tuned by perturbations such as magnetic field, pressure, and doping \cite{Canfield2016,Canfield2020}. 
         As part of our research on fragile magnets, we have been studying several fragile magnets such as itinerant ferromagnets LaCrGe$_3$ \cite{Taufour2016,Kaluarachchi2017,Gati2021,Rana2019,Rana2021} and La$_5$Co$_2$Ge$_3$ \cite{Xiang2021} which were found to exhibit avoided quantum critical points.
It is now generally believed that the ferromagnetic  quantum critical points in clean itinerant ferromagnets are avoided with some exceptions with non-centrosymmetric metals having strong spin-orbit interaction \cite{Kirkpatrick2020}.  
    Unconventional properties near quantum critical points for weak itinerant antoferromagnets have also been of particular interest, as have been studied in many compounds such as Fe-based superconductors, heavy fermions and so on.

\begin{figure*}
\centering
\includegraphics[width=2\columnwidth]{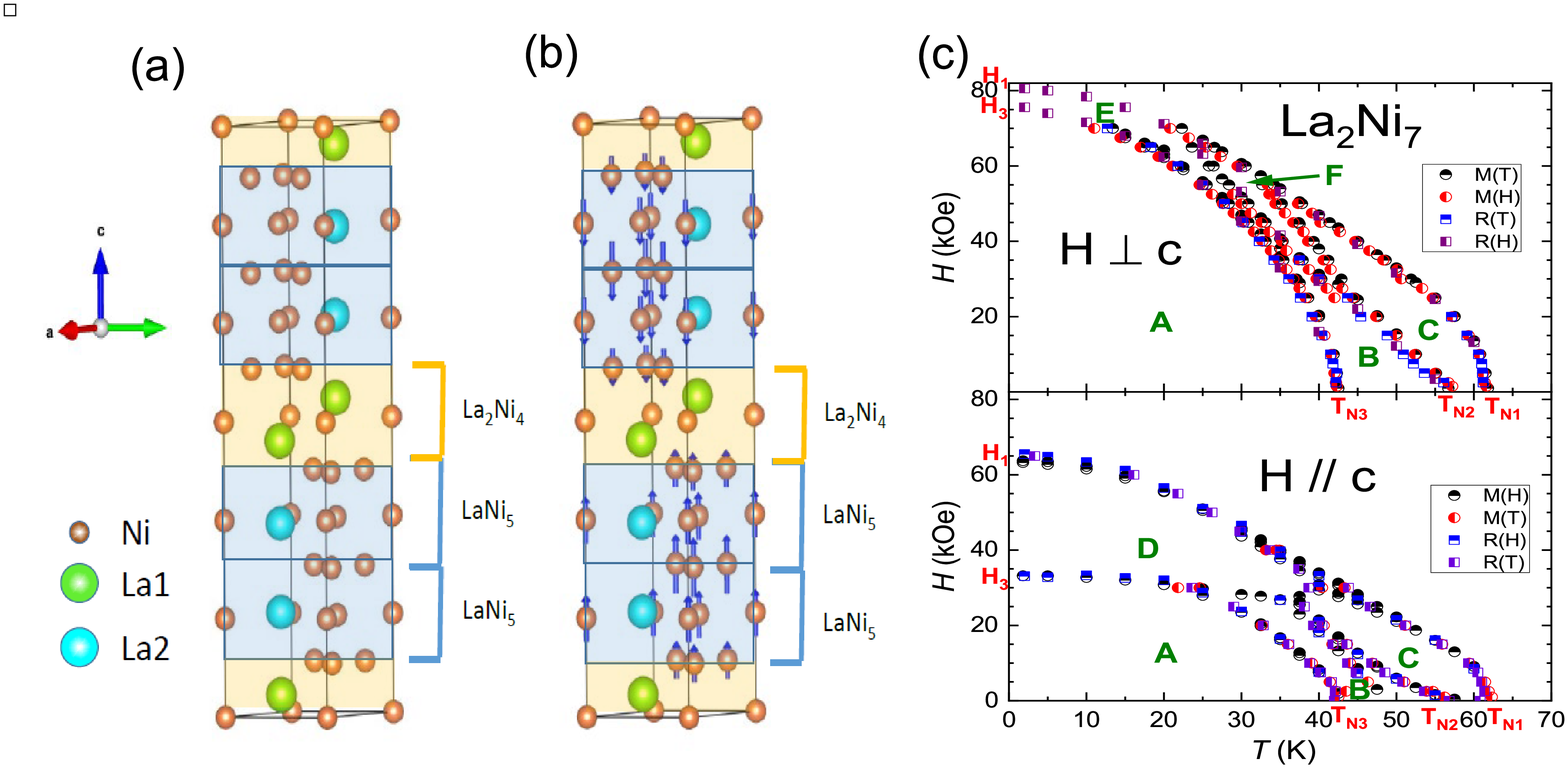}
\caption{\label{fig:Fig1} (a) Crystal structure of La$_2$Ni$_7$  (space group $P6_3/mmc$) which consists of La$_2$Ni$_4$ and LaNi$_5$ sub-units with two different La sites (La1 and La2 in the La$_2$Ni$_4$ and the LaNi$_5$ sub-units, respectively).
(b) Schematic view of spin distribution in the antiferromagnetic state (A phase) proposed by  DFT calculations \cite{Crivello2020} and  ND measurements \cite{Wilde2022}. 
(c) $H - T$ phase diagrams for $H~||~c$ (bottom) and $H \perp c$ (top) reported by Ribeiro {\it et al.} based on the $H$ and $T$ dependences of magnetization [$M(H)$ and $M(T)$] and resistivity [$R(H)$ and $R(T)$] measurements \cite{Ribeiro2022}. 
}
\end{figure*}

    Recently single crystals of the fragile itinerant antiferromagnet La$_2$Ni$_7$ with a small ordered moment were successfully synthesized \cite{Ribeiro2022}, which  makes it possible to study the physical properties of the material in detail \cite{Ribeiro2022, Wilde2022,Lee2022}. 
     La$_2$Ni$_7$ crystallizes in a Ce$_2$Ni$_7$ type hexagonal structure (space group $P6_3/mmc$) \cite{Virkar1969, Buschow1970} and the lattice parameters are $a$ = $b$ = 5.06235(11)~\AA~ and $c$ = 24.6908(8)~\AA~at room temperature \cite{Ribeiro2022}.
    As shown in Fig.~\ref{fig:Fig1}(a) \cite{VESTA}, the structure consists of La$_2$Ni$_4$ and LaNi$_5$ sub-units and the unit cell contains two blocks of [La$_2$Ni$_4$ + 2 LaNi$_5$] where two different La sites (La1 and La2 in the La$_2$Ni$_4$ and the LaNi$_5$ sub-units of the La$_2$Ni$_7$ unit cell, respectively) and five different sites for Ni atoms exist \cite{Boncour2020,Crivello2020}.
    The magnetic properties of the compound La$_2$Ni$_7$ have been investigated  since the 1980's, but only in polycrystalline form.
     Magnetic susceptibility $\chi (T)$ measurements on La$_2$Ni$_7$ in polycrystalline samples show  a maximum around 51-54~K corresponding to an antiferromagnetic (AFM) phase transition \cite{Buschow1983,Parker1983,Tuzuke1993}.
   The effective moment and Curie-Weiss temperature estimated from the $\chi (T)$ measurement  in the paramagnetic state were reported to be 0.80 $\mu_{\rm B}$/Ni and 67 K, respectively \cite{Fukase2000}.
   The saturation moment of Ni ions in the magnetically ordered state was reported to be $\sim$0.08 $\mu_{\rm B}$/Ni from magnetization curves \cite{Fukase1999}, evidencing the small ordered moments in La$_2$Ni$_7$.
   The magnetization curves  show complex temperature-dependent metamagnetic behaviors, suggesting the existence of several magnetic phases \cite{Fukase2000,Tuzuke2004}.

    In fact, from the recent detailed specific heat, magnetization, and electrical resistivity measurements on single crystals,  Ribeiro  {\it et al.} \cite{Ribeiro2022} reported the complete, anisotropic magnetic field ($H$) - temperature ($T$) phase diagrams for the magnetic fields applied parallel and perpendicular to the crystallographic $c$ axis shown in Fig.~\ref{fig:Fig1}(c).
    In zero applied magnetic field, there are a series of antiferromagnetic (AFM) phase transitions at $T_{\rm N1}$  $\sim$ 61 K, $T_{\rm N2}$  $\sim$ 56.5 K, and $T_{\rm N3}$  $\sim$ 42 K, making three different magnetically ordered phases named C, B, and A phases for $T_{\rm N2}<T<T_{\rm N1}$, $T_{\rm N3}<T<T_{\rm N2}$, and $T<T_{\rm N3}$, respectively.  
   It is noted that in  the case of  the C phase,  a small ferromagnetic (FM) component was reported in its AFM ordering \cite{Ribeiro2022}. 

   Recent neutron diffraction (ND) measurements on single crystals indicate that the B and C phases are incommensurate AFM states, while the A phase is a commensurate AFM state \cite{Wilde2022}.  
    In the A and C phases, the ordered moments are parallel to the $c$ axis, while in the B phase, there is an additional component of the ordered moments perpendicular to the $c$ axis. 
      The commensurate AFM state in the lowest temperature A phase reported by the ND measurements is consistent with the theoretical predictions \cite{Crivello2020} which suggest that two ferromagnetic unit blocks with opposite Ni spin directions are separated by the non-magnetic Ni layers in the La$_2$Ni$_4$ units [see Fig. \ref{fig:Fig1}(b)]. 
    The theoretical studies \cite{Crivello2020} also proposed that the Ni ordered moments point along the $c$ axis and their  magnitude depends on the Ni positions. As shown in Fig. \ref{fig:Fig1}(b), the magnitude is the minimum in the  La$_2$Ni$_4$ sub-units and is the maximum at the middle of the two LaNi$_5$ sub-units \cite{Crivello2020}.

  Although detailed studies of the thermodynamic, transport, and structural measurements on the itinerant antiferromagnet La$_2$Ni$_7$ using a single crystal have been performed,  it is important to investigate  the static and dynamic magnetic properties of  the fragile antiferromagnet La$_2$Ni$_7$ with the small ordered moment from a microscopic point of view.
   Nuclear magnetic resonance (NMR) and nuclear quadrupole resonance (NQR)  are powerful techniques to investigate the magnetic properties of materials from a microscopic point of view \cite{NMR_ref}.
   In particular, one can obtain the direct and local information of the magnetic state at nuclear sites, providing important information of  the magnetic structure of magnetic systems.
   In addition, the nuclear spin-lattice relaxation time, $T_1$, provides important information about magnetic fluctuations. 

   Here we report the results of $^{139}$La NMR and NQR measurements on La$_2$Ni$_7$ single crystals and oriented powder samples.
    Two distinct La signals with two  different values of quadrupolar frequencies of $\nu_{\rm Q}$ = 0.70(2) and  3.60(3) MHz are observed.
   Those are assigned to La1 and La2, respectively, based on density functional theory (DFT) calculations.
In the antiferromagnetic state of the lowest temperature A phase, the internal magnetic induction $B_{\rm int}$ at La2 was  determined to be --0.79 T at 4.3 K by the observation of the splitting of NQR lines.
     In addition, using a single crystal,  the direction of the $B_{\rm int}$ is determined to be parallel to the $c$ axis from further splittings of NQR lines under small magnetic fields.
     From NMR measurements using oriented powder sample, we found the $B_{\rm int}$ values at La1 and La2 are close to --(0.32-0.39) T and --(0.74-0.76) T at 4.3 K, respectively, in the antiferromagnetic A phase.
    This result indicates that the Ni-ordered moments are not uniform and the average magnitude of Ni-ordered moments around La2 is greater than that around La1, which is consistent with the ND results and the DFT calculations.
   In further measurements, in the saturated paramagnetic state  under a high magnetic field of 8 T parallel to the $c$ axis at 4.3 K, $B_{\rm int}$ at La1 is found to increase to --0.52 T while $B_{\rm int}$ at La2 decreases to --0.40 T.
    These results suggest that the Ni-ordered moments are more uniformly distributed at Ni positions in the saturated paramagnetic state.
    The temperature dependence of $B_{\rm int}$ at the La2 site was well reproduced by the localized moment model rather than the self-consistent renormalization (SCR) theory for weak itinerant antiferromagnets.
The temperature dependence of 1/$T_1$ measured at zero magnetic field in the paramagnetic state above $T_{\rm N1}$ can be reproduced by both the SCR theory and the localized moment model.
   These results suggest that the itinerant antiferromagnet La$_2$Ni$_7$ has magnetic properties characterized by localized moment nature,  as well as the nature of weak itinerant antiferromagnets as seen by the small Ni-ordered moments in La$_2$Ni$_7$.

 \section{Experiment}
  Hexagonal-shaped plate-like La$_2$Ni$_7$ single crystals were grown from high-temperature solutions as detailed in Ref. \cite{Ribeiro2022}.
The crystalline $c$ axis is perpendicular to the plane.
NMR and NQR measurements of $^{139}$La nuclei ($I$ = $7/2$, $\gamma_{\rm N}/2\pi$ = 6.0146 MHz/T, $Q=$ 0.21 barns) were carried out using a lab-built phase-coherent spin-echo pulse spectrometer.
The $^{139}$La-NMR spectra were obtained by recording the integrated spin-echo intensity with changing $H$ at a fixed frequency or changing resonance frequency at a constant magnetic field.
The $^{139}$La-NQR spectra in zero magnetic field were measured either by measuring the intensity of the Hahn spin echo in steps of frequency or performing the Fourier transform (FT) of the spin echo signals.
For most of NMR/NQR measurements, we performed the measurements using a powdered sample crushed from the single crystals, as the intensities of NMR/NQR signals for the single crystals are very weak, making the measurements difficult at higher temperatures.
For NMR measurements, we used a loosely packed powder sample and found that the grains of the loosely packed sample were oriented under magnetic fields and the orientation direction depends on temperature and magnetic field as shown below.
Only for NQR measurements at the lowest temperature of $T$ = 1.6 K under small magnetic fields,  a single crystal was used to determine the internal field direction.

The $^{139}$La spin-lattice relaxation rates (1/$T_{\rm 1}$) were measured using a saturation recovery method.
$1/T_1$ at each temperature for an NQR line (corresponding to a transition line for $m$ = $\pm$5/2 $\leftrightarrow$ $\pm$7/2) was determined by fitting the nuclear magnetization ($M$) versus time ($t$) using the exponential function $1-M(t)/M(\infty) = 0.214e^{-3t/T_1}+0.649e^{-10t/T_1}+0.136e^{-21t/T_1}$, where $M(t)$ and $M(\infty)$ are the nuclear magnetization at time $t$ after the saturation and the equilibrium nuclear magnetization at $t$ $\rightarrow$ $\infty$, respectively \cite{PQR}.

\section{Results and discussion}
\subsection{ $^{139}$La NMR in the paramagnetic state }

\begin{figure}[tb]
	\centering
	\includegraphics[width=\columnwidth]{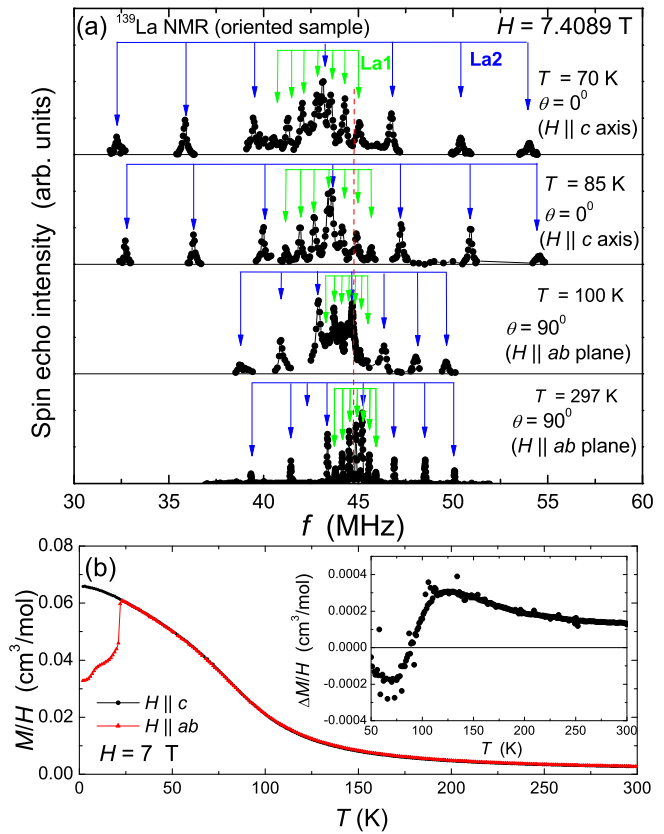} 
	\caption{(a) Frequency-swept $^{139}$La-NMR spectra under $H$~=~7.4089 T at various temperatures. 
     The green and blue lines are the position of $^{139}$La-NMR spectra calculated with the temperature independent $\nu_{\rm Q}$~=~0.70(2)~MHz for La1 and $\nu_{\rm Q}$~=~3.60(3)~MHz for La2, respectively. The vertical dashed red  line corresponds to the Larmor-frequency position. 
 (b) Temperature dependence of $M/H$ measured at $H$ = 7 T parallel to the $c$ axis [$(M/H)_c$ in black] and parallel to the $ab$ plane [$(M/H)_{ab}$ in red]. The inset shows the temperature dependence of $\Delta$$M/H$ = $(M/H)_{ab}$ $-$ $(M/H)_{c}$.}  
	\label{fig:Fig2}
\end{figure}

     Figure ~\ref{fig:Fig2}(a) shows the temperature ($T$) dependence of frequency-swept $^{139}$La-NMR spectra for the loosely packed powder sample under a magnetic field of $H$~=~7.4089 T.
       The typical spectrum for a nucleus with spin $I=7/2$ with Zeeman and quadrupolar interactions can be described by a nuclear spin Hamiltonian \cite{Slichter_book} 
\begin{eqnarray}
\centering
{\cal H} &=& -\gamma_{\rm N}\hbar(1+K) {\bf H} \cdot {\bf I} + \frac{h\nu_{\rm Q}}{6}(3I_Z^2-I^2  +\frac{1}{2}\eta(I_+^2 +I_-^2)),
\label{eq:1}
\end{eqnarray} 
where $H$ is external field, $\hbar$ is Planck's constant ($h$) divided by 2$\pi$,  and $K$ represents the NMR shift.  
The first and second terms represent Zeeman and quadrupolar interactions, respectively \cite{Hamiltonian}.  
   The nuclear  quadrupole frequency for $I=7/2$ nuclei is given by $\nu_{\rm Q} = eQV_{\rm ZZ}/14h$, where $Q$ is the nuclear quadrupole moment and $V_{\rm ZZ}$ is the maximum electric field gradient (EFG) at the La sites.
  $\eta$ is the asymmetry parameter of EFG  defined by  $\frac{V_{\rm XX} -V_{\rm YY}}{V_{\rm ZZ}}$ with $|V_{\rm ZZ}|$$\geq$$|V_{\rm YY}|$$\geq$$|V_{\rm XX}|$. 
The local symmetries at La1 and La2 are 3m. in La$_2$Ni$_7$ \cite{Virkar1969, Buschow1970}. This means that there is a threefold rotational symmetry around the $c$ axis for both La sites. Therefore $\eta$ = 0  and $V_{\rm ZZ}$ is parallel to the $c$ axis for both La sites in La$_2$Ni$_7$. Our DFT calculations also confirm this, as described below.

    In first-order perturbation theory, when the Zeeman interaction is much greater than the quadrupole interaction, one has the nuclear energy levels for $I$ = 7/2 
\begin{eqnarray}
\centering
E_m  = -\gamma_{\rm N}\hbar(1+K)Hm - \frac{h\nu_{\rm Q}}{12}(3\cos^2\theta -1 )(3m^2-\frac{63}{4}), 
\label{eq:2}
\end{eqnarray}
where  $\theta$ is the angle between the quantization axis of La nuclear spin due to the Zeeman interaction and the principal axis of the EFG. 
    Thus $^{139}$La  NMR spectrum is composed of a central transition line  and  three pairs of satellite lines shifted from the central transition line by $\pm\frac{1}{2} \nu_{\rm Q}(3\cos^2\theta -1 )$ (for the transitions of $m$ = 3/2 $\leftrightarrow$ 1/2 and -3/2 $\leftrightarrow$ -1/2), $\pm \nu_{\rm Q}(3\cos^2\theta -1 )$ (for $m$ = 5/2 $\leftrightarrow$ 3/2 and -5/2 $\leftrightarrow$ -3/2), and $\pm\frac{3}{2}  \nu_{\rm Q}(3\cos^2\theta -1 )$ (for $m$ = 7/2 $\leftrightarrow$ 5/2 and -7/2 $\leftrightarrow$ -5/2).
   It is noted that the spacing between the lines of the spectrum  for $\theta$ = 0  is twice  that for $\theta$ = 90$^{\circ}$, producing the spectrum for  $\theta$ = 0 almost two times wider than for $\theta$ = 90$^{\circ}$.

    As shown in Fig. \ref{fig:Fig1}(a), La$_2$Ni$_7$ has two inequivalent La sites: La1 and La2 belonging to the La$_2$Ni$_4$ and the LaNi$_5$ sub-units, respectively.
   The observed spectra were well reproduced by the simulated spectra [green and blue lines in Fig. \ref{fig:Fig2}(a)] from the simple nuclear spin Hamiltonian with $\nu_{\rm Q}$ = 0.70(2)~MHz for green lines and $\nu_{\rm Q}$ = 3.60(3)~MHz for blue lines.
 Here we calculated the spectrum by fully diagonalizing the nuclear spin Hamiltonian [Eq. (\ref{eq:1})] to ensure that the effects of quadrupolar interaction were fully accounted for in the determination of the NMR shifts.

    In order to assign the La sites, we have calculated the electric field gradient at each La site by a full potential linear augmented plane wave (FLAPW) method \cite{FLAPW} with a generalized gradient approximation \cite{GGA} using the lattice parameters described in Introduction.
    The $\nu_{\rm Q}$ values were calculated to be 0.50 MHz and 4.1 MHz for La1 and La2, respectively. We also found from the calculations that $V_{ZZ}$ is parallel to the $c$ axis and  the $\eta$ is zero for both sites, as expected from the local symmetries of the La sites.
    The relatively large difference in the $\nu_{\rm Q}$ values for those two La sites is consistent with the experimental results.
   Therefore we assigned the signals with $\nu_{\rm Q}$ = 3.6~MHz to La2 and those with $\nu_{\rm Q}$ = 0.7~MHz to La1.

   As shown at the bottom in Fig. \ref{fig:Fig2}(a), the observed spectrum  at $T$  = 297 K is relatively sharp and is not so-called powder pattern. 
   Here we found that  the small grains of the loosely packed powder sample were oriented due to $H$. 
  The spectrum was reproduced by the simulated spectrum with $\theta$ = 90$^{\circ}$.
   A similar but slightly broader spectrum with $\theta$ = 90$^{\circ}$ was observed $T$ = 100 K.  
    Since $V_{ZZ}$ is parallel to the $c$ axis, $\theta$ = 90$^{\circ}$ indicates that the $c$ axis of the small grains in the loosely packed sample is aligned perpendicular to the external magnetic field.
     To orient this direction, the magnetization parallel to the $ab$ plane ($M_{ab}$) must be  greater than $M$ parallel to the $c$ axis ($M_c$) under $H$ = 7.4089 T above 100 K.

  To check this, we measured the temperature dependences of $(M/H)_{ab}$ and $(M/H)_c$ under a magnetic field of $H$ = 7.0 T whose results are shown in Fig. \ref{fig:Fig2}(b) \cite{SQUID}.
   Although  the  $(M/H)_{ab}$ and $(M/H)_c$ are nearly the same in the paramagnetic state in the scale of the figure, one can see a small difference between $(M/H)_{ab}$ and $(M/H)_c$ as shown in the inset of  Fig. \ref{fig:Fig2}(b) where $\Delta$$M/H$ = $(M/H)_{ab}$ $-$ $(M/H)_{c}$ is plotted as a function of temperature. 
   $\Delta$$M/H$ changes the sign around 90 K where $(M/H)_{ab}$ is  greater than $(M/H)_{c}$ above 90 K while $(M/H)_{ab}$ is less than $(M/H)_{c}$ below that temperature. 
     The positive value of $\Delta$$M/H$ above 100 K is consistent with the NMR results.
     Therefore, our NMR data show that $\Delta$$M/H$ is enough to orient the grains of the powder sample even though it is relatively small. 
     The rotation of the grains of the powder sample is due to a torque acting on each grain, and the condition for getting an oriented powder sample depends on the size of each grain, $\Delta$$M/H$, and so on. 
Therefore, we do not discuss quantitatively here.

\begin{figure}[tb]
	\centering
	\includegraphics[width=\columnwidth]{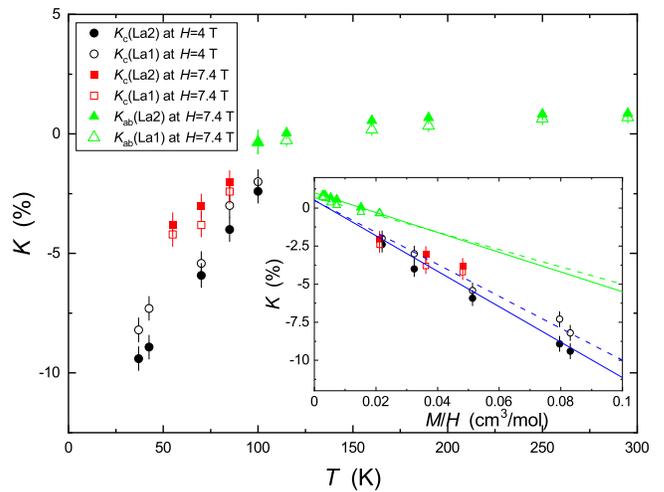} 
	\caption{Temperature dependences of the $^{139}$La-NMR shifts: $^{139}K_{ab}$ (green triangles)  and $^{139}K_{c}$ (red squares)  at $H$ = 7.4089  T. $^{139}K_{c}$ measured at $H$ =  4 T are also plotted by black symbols  \cite{M_4T}.  
    Open and filled symbols correspond to La1 and La2, respectively.  
   The inset shows the  $K$ vs. $M/H$ plots for the corresponding $H$. 
   Here we used $M/H$ measured at  $H$ = 7.0 T for the $K$-$M/H$ plot since no $M/H$ data at 7.4089 T are available \cite{SQUID}.  
   The $M/H$ data at $H$ = 4.0 T are from Ref. \cite{M_4T}.    
The blue solid and dashed lines are linear fits for the $K_c$ data of La1 and La2, respectively. 
The green solid and dotted lines are linear fits for the $K_{ab}$ data of La1 and La2, respectively. 
}
\label{fig:Fig3}
\end{figure}

 With decreasing temperature, each line becomes broader and the spacings of lines in each  NMR spectrum below $\sim$100 K become twice as those in the spectra above that temperature as typically shown in the upper two panels of Fig. \ref{fig:Fig2}(a). 
   Those spectra measured at 85 K and 70 K are well reproduced with $\theta$ = 0$^{\circ}$ with the same $\nu_{\rm Q}$ values for the two La sites (the $T$ independent $\nu_{\rm Q}$ is confirmed by NQR measurements shown below).   
    This means that the $c$ axis of the small grains of the powder sample is rotated by 90 degrees and now is oriented along the external magnetic field in that temperature range where  $M_c$  must be greater than $M_{ab}$.
  Again this is consistent with the negative values of $\Delta$$M/H$ below 90 K where $(M/H)_{c}$ is greater than $(M/H)_{ab}$ as seen in the inset of Fig. \ref{fig:Fig2}(b). 
    
   Figure \ref{fig:Fig3} shows the temperature dependence of NMR shifts determined by  fitting the observed spectra. 
The filled and open symbols correspond to La1 and La2, respectively.  
  The NMR shift data above 100 K correspond to the case for  $H \parallel ab$ plane ($K_{ab}$) while  $H\parallel c$ axis  ($K_{c}$) below 100 K at $H$ = 7.4089 T. 
   With decreasing temperature, the values of $K$ for La1 and La2 decrease similarly.
 It is noted that the change in the slopes around 100 K is mainly due to the different values of the hyperfine coupling constants for the different magnetic field directions as described below.
   To see the effect of $H$ on NMR shifts (or the magnetization),  we also measured NMR spectra at a different magnetic field of $H$ = 4.0 T whose results are shown by the black symbols in Fig. \ref{fig:Fig3}.
   As can be seen, the NMR shifts under  4.0 T are greater than those at 7.4089 T in magnitude.
    These observations are consistent with the $T$ dependence of $M/H$ under the different magnetic field where the $M/H$ is suppressed with increasing $H$ in the $T$ range of $T$ = 50 - 80 K \cite{Ribeiro2022}.
It is noted that we were able to measure $K_c$ at 100 K under $H$ = 4.0 T because $(M/H)_c$ is greater than $(M/H)_{ab}$ at 100 K at $H$ = 4.0~T \cite{M_4T}.
 
    The NMR shift consists of temperature  dependent spin shift $K_{\rm s}(T)$ and $T$ independent shift $K_{\rm 0}$; $K(T)$ = $K_{\rm s}(T)$ + $K_{\rm 0}$ where $K_{\rm s}(T)$ is proportional to the magnetization $M(T)$ divided by $H$,  $M(T)/H$,  via hyperfine coupling constant $A$: $K_{\rm s}(T)$  = $\frac{A}{\mu_{\rm B} N_{\rm A}}\frac{M(T)}{H}$.  
   Here  $N_{\rm A}$ is Avogadro's number.   
  Utilizing the $T$ dependence of $M/H$ under the corresponding magnetic fields for the  NMR measurements, we plot all $K_{\rm c}$ data as a function of $M/H$ in the inset of Fig. \ref{fig:Fig3}. 
   Here, for $K_{\rm c}$ at $H$ = 7.4089 T,   we used $M/H$ data measured at 7.0 T since no $M/H$ data at 7.4089 T are available \cite{SQUID}.  
   The blue dashed and solid lines are linear fits for La1 and La2 respectively.
    From the slopes of the blue lines, the hyperfine coupling constants for $H$ $||$ $c$  are estimated to be $A_{c}$ = --4.1(6)  and --4.5(5) T/$\mu_{\rm B}$ for La1 and La2, respectively. 
   It is noted that  the coupling constants estimated here are considered to be the total hyperfine coupling constant produced by the nearest neighboring (NN) Ni ions, as both La1 and La2 are surrounded by six NN Ni ions \cite{NN}. 
   The values of $K_0$ for $H$ $||$ $c$ are estimated to 0.5(2) \% for La1 and 0.5(2) \% for La2.
  
 We also plotted the $K_{ab}$ data (green symbols) as a function of ($M/H$)$_{ab}$ at $H$ = 7.0 T  in the inset of Fig. \ref{fig:Fig3}. 
  The green dashed and solid lines are linear fits for La1 and La2, respectively. 
From the slopes of the linear fits,  we obtained  $A_{ab}$ =  --2.2(4) T/$\mu_{\rm B}$ for La1 and --2.5(2) T/$\mu_{\rm B}$ for La2 under $H$ $||$ $ab$.
  The $K_0$ values are also estimated to be 1.0(1) and 0.7(1) \% for La1 and La2, respectively.



  \subsection{$^{139}$La NMR spectrum in magnetically ordered states}
 
\begin{figure}[tb]
	\centering
	\includegraphics[width=\columnwidth]{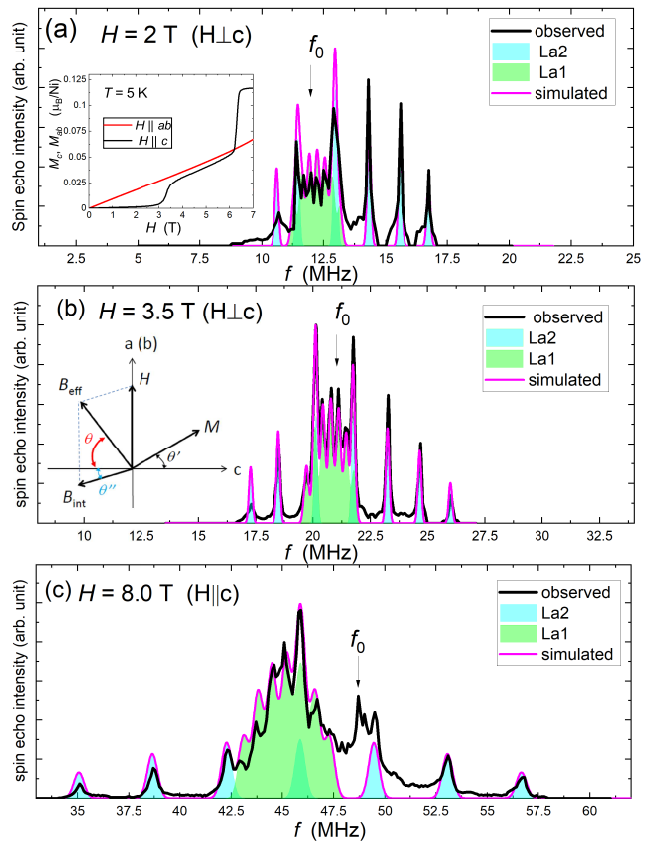} 
	\caption{Frequency-swept $^{139}$La-NMR spectra at $T$ = 4.3 K  on the loosely packed powder sample under  magnetic fields of (a) 2.0 T, (b) 3.5 T,  and (c) 8.0 T.  
     The green and blue curves are simulated spectra with $\nu_{\rm Q}$~=~0.7~MHz for La1 and $\nu_{\rm Q}$~=~3.6~MHz for La2, respectively, and different internal magnetic inductions $B_{\rm int}$ discussed in the text.
   The magenta curves are the sum of the two simulated spectra. 
   The down arrows ($f_0$)  correspond to the Larmor frequencies  for each magnetic field.
The peak around $f_0$ clearly observed in the spectrum at $H$~=~8~T probably comes from impurities \cite{impurity}.
 The inset of (a) shows the external magnetic field dependence of $M$ for $H$ $||$ $ab$ ($M_{ab}$ in red) and $H$ $||$ $c$ ($M_c$ in black) measured at 5 K from Ref. \cite{Ribeiro2022}. 
 The inset of (b) shows the schematic view of the configuration for the canting angles $\theta^{\prime}$ for $M$ and $\theta^{\prime\prime}$ for $B_{\rm int}$ from the $c$ axis,  respectively.  $\theta$ is the angle between the quantization axis of La nuclear spin ($B_{\rm eff}$)  and the $c$ axis. }
	\label{fig:Fig7}
\end{figure} 

   Figures \ref{fig:Fig7}(a)-(c) show the frequency-swept $^{139}$La-NMR spectra in magnetically ordered states at $T$ = 4.3 K for the loosely packed powder sample under magnetic fields of (a) 2.0 T, (b) 3.5 T, and (c) 8.0 T.
    As observed in the paramagnetic state, the spectra are the sum of the two sets of La NMR lines originating from La1 and La2 with $\nu_{\rm Q}$ = 0.7 MHz and 3.6 MHz, respectively.
   Since the observed spectra are relatively sharp and not so-called powder patterns, the small grains of the powder sample are also found to orient due to $H$, similar to the case of the NMR spectra in the paramagnetic state discussed in III. A.

 At $H$ = 2 T, the observed spectrum was reproduced by a sum of the two spectra from La1 (green hatched area) and La2 (light blue hatched areas) with the $H$ $\perp$ $c$ configuration. 
  This indicates that the $c$ axis of the small grains is oriented perpendicular to $H$.
This is due to the fact that the $M_{ab}$ is greater than the $M_c$ in the AFM state (A phase) at 2.0 T,  as shown in the inset of Fig. \ref{fig:Fig7}(a) where the $H$ dependences of the $M_c$ and $M_{ab}$ at $T$ = 5 K are shown (the data are from Ref. \cite{Ribeiro2022}).
   In the simulations of the spectra, we introduced the internal magnetic inductions ($B_{\rm int}$) at the La sites  which are produced by the Ni ordered moments. 
  This is reasonable since the Ni moments are ordered in the AFM state.

 For La2,  we used  $B_{\rm int}^c$ =  --0.76(1) T  parallel to the $c$ axis and $B_{\rm int}^{ab}$ = --0.03(1) T perpendicular to the $c$ axis. 
   At zero magnetic field, as described in the next subsection III. C, the direction of the  $B_{\rm int}$ at La2 is determined to be parallel to the $c$ axis (that is, parallel to the Ni-ordered moments) in the A phase.  
     Under a magnetic field perpendicular to the $c$ axis, however, the $B_{\rm int}$ is no longer parallel to the $c$ axis  because of spin canting as shown in the inset of Fig. \ref{fig:Fig7}(b). 
     This is the reason why we needed to introduce the  $B_{\rm int}^{ab}$ in the simulation. 
    The canting angle $\theta^{\prime}$ of the Ni ordered moment from the $c$ axis can be calculated from magnetization data  since $\theta^{\prime}$ is given by sin$^{-1}$($M_{ab}/M_{\rm s})$ where $M_{\rm s}$ is the saturation of magnetization. 
   Taking the saturated value of $M_c$ at 7.0 T as $M_s$ [see the inset of Fig. \ref{fig:Fig7}(a)],  $\theta^{\prime}$ is estimated to be 8.9$^{\circ}$ at $H$ = 2.0 T. 
    It is noted that the angle  $\theta^{\prime}$ is not the same as $\theta^{\prime\prime}$ because of the anisotropy in the hyperfine coupling constants. 
   Here the  $\theta^{\prime\prime}$ is the angle between $B_{\rm int}$ and the $c$ axis. 
    From $B_{\rm int}^c$ =  --0.76(1) T and $B_{\rm int}^{ab}$ = --0.03(1) T, $\theta^{\prime\prime}$ is estimated to be 2.3(5)$^{\circ}$.
    This value is smaller than  $\theta^{\prime\prime}$  = 4.9(9)$^{\circ}$  which is estimated from the $M_{ab}$/$M_s$ at 2 T  by using  $A_{c}$ = --4.5(5) T/$\mu_{\rm B}$  and $A_{ab}$ =  --2.5(2) T/$\mu_{\rm B}$.
  As shown in the inset, the effective magnetic induction  at the La site ($\bf{B}_{\rm eff}$) given by the vector sum of $\bf{B}_{\rm int}$ and $\bf{H}$ is not perpendicular to the $c$ axis due to the $B_{\rm int}$.   
    We also estimated the angle $\theta$ between the quantization axis of La nucleus and the $c$ axis. 
Using the experimentally obtained $B_{\rm int}$, $\theta$ is estimated to be  68(2)$^{\circ}$.  

  For the simulation of spectrum for La1, $B_{\rm int}^c$ =  --0.39(1) T  and $B_{\rm int}^{ab}$ = --0.015(5) T were used. 
$\theta^{\prime\prime}$ and $\theta$ are estimated to be 2.2(4)$^{\circ}$ and 79(2)$^{\circ}$, respectively. 
$\theta^{\prime\prime}$ is estimated to be 4.8(5)$^{\circ}$ from the $M$ data utilizing the hyperfine coupling constants.

The observed spectrum at $H$ = 3.5 T  is also reproduced with the $H$ $\perp$ $c$ configuration and a similar set of the parameters of $B_{\rm int}^c$ = --0.32(2) [--0.74(1)] T and $B_{\rm int}^{ab}$ = --0.05 T [--0.03(1)] T for La1 [La2], respectively, with the same $\nu_{\rm Q}$ values. 
  Again $H$ $\perp$ $c$  is consistent with the $M$ data where  the $M_{ab}$ is greater than the $M_c$ at $H$ = 3.5 T [see the inset of Fig. \ref{fig:Fig7}(a)].
   At $H$ = 3.5 T, we estimate  $\theta^{\prime}$ = 15.5$^{\circ}$ from the $M$ data. 
$\theta^{\prime\prime}$ = 5.4(5)$^{\circ}$ [3.9(4)$^{\circ}$] and $\theta$ = 78(1)$^{\circ}$ [83.8(6)$^{\circ}$] for La1 [La2] are evaluated from the NMR data.
 
 When we simulated the spectra at $H$ = 2 and 3.5 T, it turned out that the changes in $B_{\rm int}^{c}$ alter the spacings between lines. 
   On the other hand, $B_{\rm int}^{ab}$ mainly shifts the whole spectrum without changing the spacing between the lines much. 
   This is because the direction of the external magnetic field is perpendicular to the $c$ axis. 
    Therefore we can determine the relatively accurate value of $B_{\rm int}^{c}$ by simulations. 
   In contrast, we were not able to determine $B_{\rm int}^{ab}$ precisely since one needs to take the contribution of Knight shift $K$ into consideration. 
   Since La$_2$Ni$_7$ is in a metallic state, we should have a finite value of $K$ along the $H$ direction. 
   This $K$ shifts the whole spectrum as $B_{\rm int}^{ab}$ does. 
   Thus, the estimates of the $B_{\rm int}^{ab}$ values depend on the value of $K$. 
   In our simulations, we used $K$ = 0.5 \% which is estimated from the value of 1/$T_1T$ in the AFM state using the Korringa relation \cite{T1T_AFM}. 
   Although we consider that $K$ = 0.5 \% is reasonable because this value is close to the values of $K_0$ estimated from the $K-\chi$ analysis, there must be relatively large uncertainty in the estimation of $B_{\rm int}^{ab}$.  
   Therefore, the estimated directions of $B_{\rm int}$ and $B_{\rm eff}$ discussed above should be considered as tentative ones.
 

   When the system is in the saturated paramagnetic state under $H$ = 8 T, the spectra become wider due to the change from the $H$ $\perp$ $c$ to $H$ $||$ $c$ configurations.
   This corresponds to the $c$ axis of the small grains of the powder sample oriented along $H$, which is again consistent with the $M$ data where $M_c$  is now greater than  $M_{ab}$ above $H$~=~6.2~T as can be seen in the inset of Fig. \ref{fig:Fig7}(a).
    By fitting the spectra with keeping the same $\nu_{\rm Q}$ values for the La sites, we found that  $B_{\rm int}^c$ = --0.52(1)~T and --0.40(1)~T for La1 and La2, respectively.

     Utilizing the hyperfine coupling constants and the internal magnetic inductions, the average values of Ni ordered moments around La1 and La2 from  $\mu$ = $\sqrt{(\frac{B_{\rm int}^c}{A_c})^2+(\frac{B_{\rm int}^{ab}}{A_{ab}})^2}$  are estimated to be 0.10(1) $\mu_{\rm B}$ for La1 and 0.17(1) $\mu_{\rm B}$ for La2 at $H$ = 2 T in the AFM state (A phase). 
Although we have uncertainty in $B_{\rm int}^{ab}$, the estimated $\mu$ values should be relatively reliable because the values of $\mu$ are mainly determined by the $(\frac{B_{\rm int}^{c}}{A_{c}})^2$ term in the formula due to   $(\frac{B_{\rm int}^{c}}{A_{c}})^2$ $>>$  $(\frac{B_{\rm int}^{ab}}{A_{ab}})^2$. 
At $H$ = 3.5 T,  the values of $\mu$  were estimated to 0.09(1) $\mu_{\rm B}$ and 0.17(1) $\mu_{\rm B}$ for La1 and La2, respectively.
   The average values of Ni-ordered moments around La1 is 56 \% of that around La2, suggesting a nonuniform distribution of the Ni-ordered moments in the AFM state. 
  This is qualitatively consistent with the AFM state predicted by the theory and the ND measurement,  where the maximum amplitude of the Ni-ordered moments is expected at the middle of two LaNi$_5$ sub-units close to La2 while the minimum amplitude of the Ni-ordered moments is proposed  at Ni ions in the La$_2$Ni$_4$ sub-unit close to La1.

    We also estimate the average values of Ni-ordered moments to be 0.12(1) $\mu_{\rm B}$ and 0.09(1) $\mu_{\rm B}$ around  La1 and La2, respectively,   in the  saturated paramagnetic state. 
    According to the theoretical calculations,  in the ferromagnetic state,  the ferromagnetic units are stacked ferrromagnetically while keeping the nonuniform distribution of the moments as in the AFM state \cite{Crivello2020}.
    In this case, we expect almost no change in the average values of Ni-ordered moments around both La sites in the  saturated paramagnetic state as compared to the AFM state. 
   This is inconsistent with the NMR results showing that the distribution of the average values of Ni-ordered moments in the saturated paramagnetic state is different from the case of AFM state.

  It is noted that the NMR spectrum for La2 at $H$  = 2 T shows 6 lines instead of the expected 7 lines. 
This is because one of the NMR lines overlaps with another line due to the relatively large $B_{\rm int}^c$ = --0.75 T along the $c$  axis which is perpendicular to the external magnetic field of 2 T. 
   Qualitatively this could be understood by taking two effects into consideration: (1) the change in the angle $\theta$  between the quantization axis of the nuclear spin and $V_{zz}$ and (2) the increase of the effective magnetic induction due to the $B_{\rm int}$. 
   Since the position of the satellite lines changes following 3cos$^2\theta$$-$1,  each satellite line will shift toward the central transition line with increasing $B_{\rm int}$. At the same time, the effective magnetic induction increases, making the NMR lines shift to a higher frequency. By combing the two effects, one could qualitatively understand the spectrum at $H$ = 2 T. 
   As shown, we were able to reproduce the observed spectrum by the simulation where the nuclear spin Hamiltonian including the internal magnetic induction was fully diagonalized. 

\subsection{$^{139}$La NQR spectrum}

\begin{figure}[tb]
	\centering
	\includegraphics[width=\columnwidth]{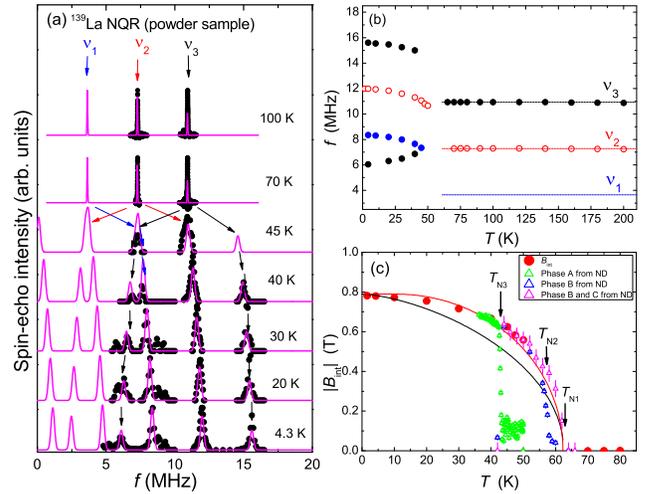} 
	\caption{(a) $^{139}$La NQR spectra for La2 at zero magnetic field at various temperatures in both the paramagnetic and magnetically ordered states. The magenta curves  are simulated spectra.   
     (b) Temperature dependence of the resonance frequency for each peak. The symbols in black, red and blue represent for $\nu_3$, $\nu_2$ and $\nu_1$ lines, respectively.  
    (c) Temperature dependence of the magnitude of the internal magnetic induction ($|B_{\rm int}|$) at La2. The values of $B_{\rm int}$  are estimated by fitting the the spectra, and the simulated spectra are shown by curves in magenta in (a).  
   The expected temperature dependences of $B_{\rm int}$ based on the ND data \cite{Wilde2022}  are also plotted in (c) by the green open triangles (the $c$ components of the ordered moments in the commensurate antiferromagnetic A phase), the blue open triangles  (the $ab$ plane components of the ordered moments in  the incommensurate antiferromagnetic B phase), and the magenta open triangles (the $c$  components of the ordered moments in the incommensurate antiferromagnetic  B and C phases).  
    The ND data sets  shown by the green, blue, and magenta symbols were normalized to scale the NMR data at 40 K,  45 K, and 50 K, respectively. 
     The black and red curves show the expected temperature dependences of $B_{\rm int}$ based on the SCR theory and the localized moment picture, respectively (see text). 
}
	\label{fig:Fig4}
\end{figure} 

   At zero magnetic field, the electric quadrupole Hamiltonian (the second term of Eq. \ref{eq:1}) for the case of $I$ =7/2 produces three equally spaced NQR lines at resonance frequencies of $\nu_{\rm Q}$, 2$\nu_{\rm Q}$, and 3$\nu_{\rm Q}$ corresponding to the transitions of $m$ = $\pm$1/2 $\leftrightarrow$ $\pm$3/2,  $\pm$3/2 $\leftrightarrow$ $\pm$5/2 and $\pm$7/2 $\leftrightarrow$ $\pm$5/2, respectively. 
   From NMR measurements, the values of $\nu_{\rm Q}$ are estimated to be 3.60(3) MHz for La2, so one expects three NQR lines at 3.6 MHz, 7.2 MHz, and 10.8 MHz, named  $\nu_1$, $\nu_2$, and $\nu_3$ lines, respectively. 
   In fact, we observed such NQR lines in the paramagnetic state as shown in Fig. \ref{fig:Fig4}(a) where the $\nu_1$ line was not measured due to the limitation of the resonance frequency of our spectrometer (we also could not measure the NQR lines for the La1 site due to the small value of $\nu_{\rm Q}$ = 0.7 MHz).  
   The observed NQR spectra, that is, the peak frequencies for the $\nu_2$ and $\nu_3$ lines for the La2 site are well reproduced with $\nu_{\rm Q}$ = 3.63(1) MHz as shown by the calculated spectra in magenta, and those peak frequencies are independent of temperature in the paramagnetic state [see Fig. \ref{fig:Fig4}(b)], indicating no change of $\nu_{\rm Q}$ with temperature.

   In the AFM state, we found that each line splits into two lines, as shown in Fig. \ref{fig:Fig4}(a).
   This is due to the appearance of $B_{\rm int}$ at La2 produced by Ni-ordered moments in the AFM state.  
The $B_{\rm int}$ lifts the degenerate nuclear spin levels of $\pm$$m$ states.
   To estimate $B_{\rm int}$, we calculated NQR spectra using a Hamiltonian where a Zeeman term due to $B_{\rm int}$, similar to the first term of Eq. \ref{eq:1},  was added to the electrical quadrupole Hamiltonian. 
  We found that the observed spectra were well reproduced by taking $B_{\rm int}$ parallel to the $c$ axis, as shown by the calculated spectra in magenta in the magnetically ordered state. 
   The value of $|$$B_{\rm int}$$|$  is estimated to be  0.79 T at 4.3 K.
   It is noted that we also observed the small signals around 6 and 9 MHz at low temperatures below 30 K which were not reproduced by the simulation. The origin of the lines is not clear at present. It might be related to the impurity signals observed at Larmor frequency in the NMR spectrum measured at 8 T described in III. B.  
   The temperature dependence of $B_{\rm int}$ at the La2 site is shown in Fig. \ref{fig:Fig4}(c), which corresponds to the temperature dependence of sublattice magnetization $M_{\rm sub}$.
    Here we also plotted the expected temperature dependence of $M_{\rm sub}$ ($\propto$ $B_{\rm int}$) based on the ND data \cite{Wilde2022} by the green open triangles (the $c$ components of the ordered moments in the commensurate antiferromagnetic A phase), the blue open triangles  (the $ab$ plane components of the ordered moments in  the incommensurate antiferromagnetic B phase), and the magenta open triangles (the $c$  components of the ordered moments in the incommensurate antiferromagnetic  B and C phases).  
    The ND data sets shown by the green, blue, and magenta symbols were normalized to scale the NMR data at 40 K,  45 K, and 50 K, respectively. 
It is noted that the phase transitions at $T_{\rm N2}$ and $T_{\rm N1}$ are of second order while the first order phase transition was observed at $T_{\rm N3}$ \cite{Wilde2022}.
  It is also noted that the temperature dependences of $M_{\rm sub}$ from the ND measurements in the magnetically ordered states are overlapped with that of $B_{\rm int}$ determined by the NMR measurements although the overlapped temperature range ($T = 35 - 50$ K) is not wide. 

According to self-consistent renormalization (SCR) theory,  the temperature dependence of $M_{\rm sub}$  for the case of  weak itinerant antiferromagnets is given by \cite{SCR1}
\begin{eqnarray}
\centering
M_{\rm sub}(T) \propto \sqrt{1-(T/T_{\rm N})^{3/2}}
\label{eq:3}
\end{eqnarray} 
The black curve is the calculated temperature dependence of $|B_{\rm int}|$  (=$B_{\rm int,0}\sqrt{1-(T/T_{\rm N})^{3/2}}$) based on the above formula with $B_{\rm int,0}$ = 0.79 T and $T_{\rm N}$ = 62 K, which  does not reproduce the experimental data well.
 Since the phase transition at $T_{\rm N3}$ is of first order, we also tried to fit  only for the data above $T_{\rm N3}$ using the formula. 
However,  we could not fit the data well with a reasonable value of $B_{\rm int,0}$. 
In addition, it is clear that the observed temperature dependence of $B_{\rm int}$ cannot be explained by the formula using the different values of $T_{\rm N}$. 
    Thus we conclude that the temperature dependence of the  $M_{\rm sub}$ in La$_2$Ni$_7$ cannot be explained by the model of weak itinerant antiferromagnets.   
  Instead, the temperature  dependence of $B_{\rm int}$ was found to be reasonably reproduced by a  Brillouin function which was calculated based on the Weiss molecular field model shown by the red curve where we tentatively assumed $S$ = 1  and used  $T_{\rm N}$ = 62 K. 
 Here we normalized the calculated temperature dependence to scale with $|B_{\rm int}|$ at the lowest temperature.
   These results may indicate that the magnetic state of the Ni ions is more likely described by the local-moment picture, contrary to the anticipation from the small Ni-ordered moments in the itinerant antiferromagnet La$_2$Ni$_7$.
  Interestingly, similar discrepancy between $M_{\rm sub}$ and the SCR theory has been reported in weak itinerant antiferromagnets such as V$_5$Se$_8$ \cite{Kitaoka1980V5Se8}, V$_5$S$_8$ \cite{Kitaoka1980V5S8}, CrB$_2$\cite{Kitaoka1980CrB2}, and YMn$_{12}$ \cite{Yoshimura1990}.
Further theoretical and experimental studies are called for to understand the local moment nature in those weak itinerant systems. 

\begin{figure}[tb]
	\centering
	\includegraphics[width=\columnwidth]{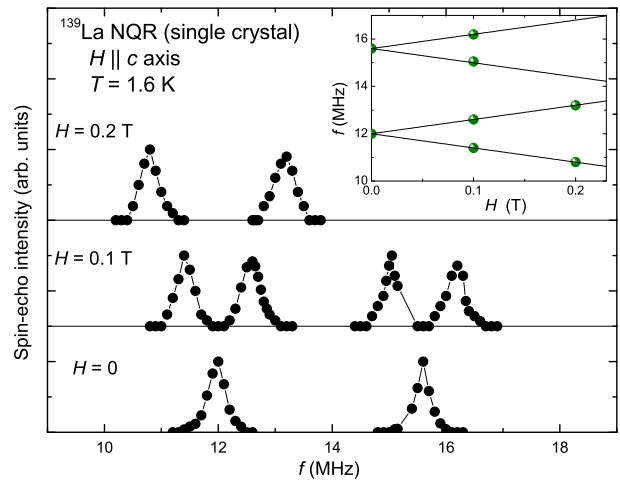} 
	\caption{$^{139}$La NQR spectra for La2 under small magnetic fields at 1.6 K. The inset shows the external magnetic field dependence of peak frequencies of the split lines for the $\nu_2$ and $\nu_3$ lines. The solid lines are the expected $H$ dependence of peak frequencies when $B_{\rm int}$ is parallel or antiparallel to the $c$ axis. 
}
	\label{fig:Fig5}
\end{figure}

The direction of $B_{\rm int}$ in the AFM state is also directly confirmed by $^{139}$La NQR spectrum measurements on the single crystal in nonzero $H$.
Figure \ref{fig:Fig5} shows the $H$ dependence of NQR spectrum at 1.6 K under small magnetic fields applied parallel to the $c$ axis using a single crystal.
With the application of small $H$ less than 0.2 T, we found that one of the split $\nu_2$ lines that is shifted to higher frequency side (around 12 MHz) further splits into two lines as shown in Fig. \ref{fig:Fig5}.
    Similar splitting is also observed for the $\nu_3$ line around 15.6 MHz under small magnetic fields.
These results clearly indicate the existence of two La2 sites with  $B_{\rm int}$ parallel or antiparallel to the $c$ axis.
As discussed above, the effective field at the La site is given by the vector sum of $\bf{B}_{\rm int}$ and $\bf{H}$, i.e., $|$$\bf{B}_{\rm eff}$$|$ = $|$$\bf{B}_{\rm int}$ + $\bf{H}$$|$, the resonance frequency is expressed for $H$ $||$ $B_{\rm int}$ as $f$ =$\frac{\gamma_{\rm N}}{2\pi}$$|B_{\rm eff}|$ = $\frac{\gamma_{\rm N}}{2\pi}$$|B_{\rm int}$ $\pm$ $H|$.
As shown in the inset, the $H$ dependence of the split peak frequencies is well reproduced by the relation using the $\frac{\gamma_{\rm N}}{2\pi}$ = 6.0146 MHz/T for La nucleus, evidencing that the $B_{\rm int}$ is parallel or antiparallel to the $c$ axis without any obvious deviation.
This result also clearly shows the appearance of two sublattices, a direct confirmation of the antiferromagnetic state.

\begin{figure}[tb]
	\centering
	\includegraphics[width=\columnwidth]{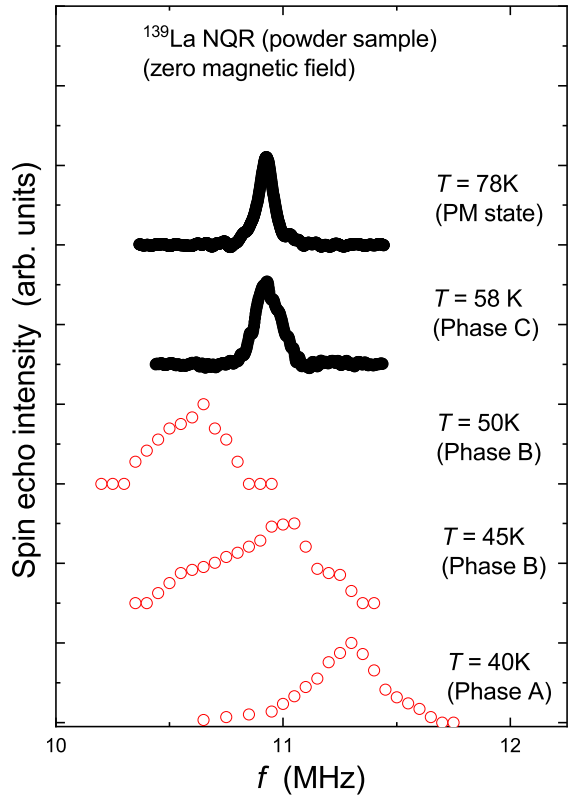} 
	\caption{$^{139}$La NQR spectra for La2 at zero magnetic field at various temperatures in both the paramagnetic and the magnetically ordered states. The spectra shown by black solid and red open circles are the $\nu_3$ and $\nu_2$ lines, respectively.
}
	\label{fig:Fig6}
\end{figure}

   To see how the NQR spectra change around magnetic phase transition temperatures for the investigation of the AFM states of the B and C phases, we measured the NQR spectra around 11 MHz at zero magnetic field for the three phases.
   Figure \ref{fig:Fig6} shows the measured NQR spectra for the powder samples in the three magnetic phases together with the one in the PM state.
    The line around 10.9 MHz in the PM state corresponds to the $\nu_3$ line.
    When the sample is in the AFM state at $T$ = 58 K (C phase), we found that the position of the NQR line does not change, although the line becomes slightly broad.
   According to the ND measurements, the C phase  is proposed as an incommensurate AFM state with the Ni-ordered moment parallel to the $c$ axis.
   Therefore the small broadening could be due to a small $B_{\rm int}$ in the phase.
   With further decrease in temperature, the signal around 10.9 MHz disappeared when the system entered the B phase as shown by the spectrum at $T$ = 50 K.
Instead, we observed the slightly broader NQR line centered at 10.5 MHz, lower than the peak frequency of 10.9 MHz in the paramagnetic state and the C phase.
   We assigned the line to be one of the split $\nu_{2}$ lines due to $B_{\rm int}$ along the $c$ axis because the position of the line smoothly connects to that of the line in the A phase [see Fig. \ref{fig:Fig4}(c)].
    This suggests that the dominant component of the $B_{\rm int}$ at La2 is parallel or antiparallel to the $c$ axis in the B phase, similar to the case in the A phase.
 According to the ND data, the propagation vectors parallel and perpendicular to the $c$ axis are incommensurate in the B phase.
This indicates that the Ni ordered moments are no longer ferromagnetic block \cite{Wilde2022}. 
  In this case,  one expects the large distribution of $B_{\rm int}$ at the La sites. 
Therefore the NMR data would be inconsistent with the ND results. 
The slightly broader lines are due to small distributions of $B_{\rm int}$.
This suggests  slight distributions of the Ni-ordered moments in the B phase, might be related to the in-plane component of the incommensurate  AFM state reported from the ND data.
   In contrast, as we discussed in the previous subsection, the commensurate  AFM state (Phase A) reported from the ND measurements is consistent with the NMR results.

   \subsection{$^{139}$La spin-lattice relaxation time  $T_1$}

\begin{figure}[tb]
	\centering
	\includegraphics[width=\columnwidth]{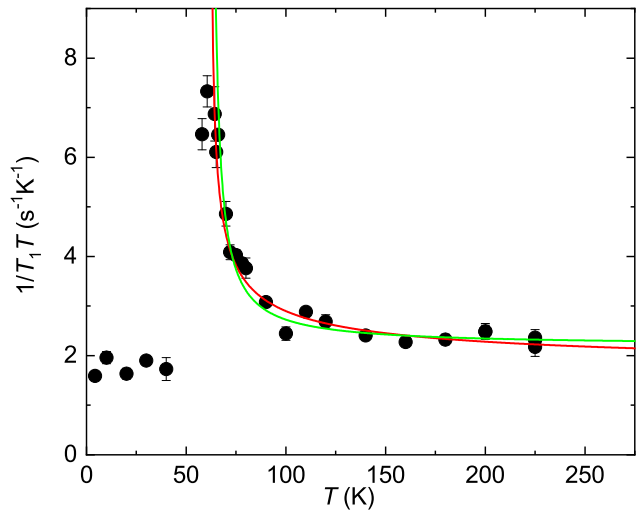} 
	\caption{Temperature dependence of 1/$T_1T$ measured at the 3$\nu_{\rm Q}$ line in the paramagnetic and magnetic states. The red and green curves are the calculated results based on the SCR theory and the localized moments picture, respectively (see text). 
 }
	\label{fig:Fig8}
\end{figure} 
    
 Figure \ref{fig:Fig8} shows the $T$ dependence of 1/$T_1T$ measured at the $\nu_{3}$ line of $^{139}$La NQR spectrum under zero magnetic field.
In the paramagnetic state, 1/$T_1T$ is nearly constant above 150 K and starts to increase below that temperature with decreasing $T$, and shows a peak around $T_{\rm N1}$ due to critical slowing down of spin fluctuation, characteristics of second-order phase transition.
In the AFM state (A phase), the 1/$T_1T$ = constant behavior expected for metals is observed  \cite{T1T_AFM}. 
This is consistent with the itinerant nature of La$_2$Ni$_7$.

Since the system has small ordered moments, we first fit the data based on the SCR theory.
For the case of weak itinerant antiferromagnets, the SCR theory predicts the temperature dependence of 1/$T_1T$ in paramagnetic states given by \cite{SCRT1,SCR2}\begin{eqnarray}
\centering
1/T_1T = \frac{a}{\sqrt{T-T_{\rm N}}} + b
\label{eq:4}
\end{eqnarray} 
where the first term originates from AFM fluctuations around a wave vector $q = Q$ ($Q$ being antiferromagnetic wave vector) and the second term is due to Korringa-type relaxation, a characteristic feature of metallic materials. 
   The red curve in the paramagnetic state shows the calculated results based on the SCR theory with a set of parameters of $a$ = 8 (s$^{-1}$K$^{-0.5}$), $b$ = 1.6 (s$^{-1}$K$^{-1}$) and $T_{\rm N}$ = 62 K, which reproduces the experimental data in contrast to the case of the $T$ dependence of $B_{\rm int}$. 

On the other hand, as the $T$ dependence of $B_{\rm int}$ suggests the localized moment nature rather than the weak itinerant antiferromagnets, we also  analyzed  the $T_1$ data with the localized moment model described by 
\begin{eqnarray}
\centering
1/T_1T = \frac{a^{\prime}}{T-\Theta} + b^{\prime}
\label{eq:5}
\end{eqnarray} 
where the first term originates from the paramagnetic uniform spin  fluctuations of localized spins described by Curie-Weiss behavior of the magnetic susceptibility, i.e., $1/T_1$ $\propto$ $\chi$$T$  and the second term is Korringa-type relaxation, a characteristic feature of metallic materials as in eq. \ref{eq:4}. 
With $a^{\prime}$ = 20 (s$^{-1}$),  $b^{\prime}$ = 2.2 (s$^{-1}$K$^{-1}$), and $\Theta$ = 62 K, the experimental data are also reproduced reasonably  as shown by the green curve in Fig. \ref{fig:Fig8}. 
    Therefore, both the SCR theory and the localized moment moment can explain the observed $T$ dependence of 1/$T_1$, although those are different in character of spin fluctuations. 
Thus, given the small Ni-ordered moments evidenced by the magnetization and present NMR spectrum measurements,  our experimental results may suggest that La$_2$Ni$_7$ has both the characteristic natures of weak itinerant and localized moment antiferromagnets.

 \section{Summary}
  
     In summary, we have performed the microscopic characterization of the itinerant antiferromagnet La$_2$Ni$_7$ by $^{139}$La NMR and NQR measurements. 
    Our results show the presence of commensurate antiferromagnetic A phase with Ni ordered moments aligned along crystalline $c$ axis. 
    Furthermore, a non-uniform distribution of the Ni-ordered moments was found with the comparative analysis of inequivalent La1 and La2 sites. 
    In the saturated paramagnetic  state under a magnetic field of 8 T at 4.3 K, the Ni-ordered moments were found to become more uniform. 
     The temperature dependence of the sublattice magnetization measured by the internal magnetic induction $B_{\rm int}$  at the La2 site is  reproduced  by the local-moment model. 
    The temperature dependence of $^{139}$La spin-lattice relaxation time ($T_1$) suggests that the paramagnetic to antiferromagnetic  phase transition at $T_{\rm N3}$ is second order in nature and the its temperature dependence in the paramagnetic state  can be produced by both the local-moment model and the self-consistent renormalization model for weak itinerant antiferromagnets. 
     Our NMR results suggest a  peculiar magnetism in the itinerant antiferromagnet La$_2$Ni$_7$ that exhibits the properties of both localized and itinerant nature.

 \section{ Acknowledgments} 
The authors would like to thank J. Wilde and A. Sapkota for providing the neutron diffraction data and also for fruitful discussions. 
The research was supported by the U.S. Department of Energy, Office of Basic Energy Sciences, Division of Materials Sciences and Engineering. Ames National Laboratory is operated for the U.S. Department of Energy by Iowa State University under Contract No.~DE-AC02-07CH11358.


\begin{thebibliography}{99}
\bibitem{Hamiltonian} Here we omitted in the quadrupole Hamiltonian a term containing the asymmetry parameter  $\eta$ of the electric field gradient since it is found to be zero from NQR experiment and also the DFT calculation.
\bibitem{SQUID} The magnetization measurements were performed with a Quantum Design Magnetic Properties Measurement System (MPMS). Due to the limitation of the magnetic field of our MPMS, the magnetization was measured at 7 T, the highest magnetic field of the MPMS.
\bibitem{NN} In addition to the 6 NN Ni ions for the La1 and La2 sites, there are 3 second NN  and  3 third NN Ni ions for the La1 site and 6 second NN  and 6 third NN Ni ions for the La2 site.
\bibitem{T1T_AFM}
Utilizing the so-called Korringa relation of $T_1TK^2$  = $\frac{\hbar}{4\pi k_{\rm B}} \left(\frac{\gamma_{\rm e}}{\gamma_{\rm N}}\right)^2$ where $\gamma_{\rm e}$ is the electron gyromagnetic ratio,  the Knight shift due to the conduction electrons is estimated to be $K$ = 0.5 \% from 1/$T_1T$ = 1.85 (sK)$^{-1}$ in the AFM state. Since we do not have $T_1$ data for La1,  the value of $K$ for La1 due to the conduction electrons is unknown. However, we used $K$ = 0.5 \% for both La sites since the $K_0$ values estimated the $K-\chi$ plot analysis are close for the La sites.
\bibitem{impurity}
The impurity signals were not clearly observed in the spectra at  $H$ = 2 and 3.5 T in Fig. 4.  We consider this is because the intensity of the intrinsic signals is higher than that of impurity signals at $f_0$ in the spectra, making the observation of the impurity signals difficult.
\end{thebibliography}
\end{document}